\documentclass[
aps,
pre,
reprint,
superscriptaddress,
nofootinbib
]{revtex4-2}

\usepackage{amsmath}
\usepackage{amssymb}
\usepackage{bm}
\usepackage{mathtools}
\usepackage{booktabs}
\usepackage{xcolor}
\usepackage{orcidlink}
\usepackage{hyperref}
\hypersetup{
    colorlinks=true,
    linkcolor=blue,
    citecolor=blue,
    urlcolor=blue
}
\usepackage{comment}
\usepackage{graphicx}
\usepackage{caption}
\usepackage{subcaption}
\usepackage{ragged2e}

\newcommand{\dd}{\mathrm{d}}

\begin{document}

\title{A Generalized Model for Disordered Random Sequential Adsorption with Charge-Dependent Deposition}

\author{G. Palacios \orcidlink{0000-0003-2609-9400}}
\email{guillermo.roque@servidor.uepb.edu.br}
\affiliation{Centro de Ciências Exatas e Sociais Aplicadas, Universidade Estadual da Paraíba, 58706-550 Patos, PB, Brazil}
\affiliation{ Departamento de Física, Universidade Federal de Pernambuco, 50670-901 Recife, PE, Brazil}
\author{A. M. S. Macêdo \orcidlink{0000-0002-4522-031X}}
\affiliation{ Departamento de Física, Universidade Federal de Pernambuco, 50670-901 Recife, PE, Brazil}

\date{\today}


\begin{abstract}
We generalize a one-dimensional random sequential adsorption model (RSA) of charged unit segments in which the deposition position follows a beta kernel determined by the charges bounding the available gap and those of the incoming particle. The interaction parameter \(0\leq\lambda<1\) interpolates between uniform car parking and strongly localized deposition. We derive a four component generating function equation that provides exact recursions for the complete occupation statistics, although only the mean and variance are analyzed in detail here. Exact finite-shell solutions serve as benchmarks for the Gauss--Jacobi quadrature and Monte Carlo simulations. We prove that, although boundary charges affect finite size behavior, the four boundary states have the same asymptotic mean and variance densities, consequently, we prove that the jammed density is self-averaging. Numerical results show that charge selectivity increases the coverage, especially for mixtures dominated by opposite endpoint charges, which also exhibit strongly reduced fluctuations near \(\lambda=1\). The model provides a tractable connection between uniform RSA, dynamically generated disorder, and interaction-driven deposition.

\end{abstract}

\maketitle

\section{Introduction}

Random sequential adsorption (RSA) is one of those nonequilibrium models whose rules are much simpler than the structures they produce. Particles are proposed one after another, accepted when they do not overlap previously deposited particles, and permanently fixed after acceptance. Nothing moves afterward, and no accepted event can be reversed. This elementary irreversibility is enough to generate spatial correlations, memory of the deposition history, and a jammed state whose density is generally lower than the densest geometrically allowed packing. Unlike an equilibrium hard-particle system, an RSA configuration cannot explore its configuration space and is not described by a Gibbs measure \cite{Widom1966,Evans1993,Talbot2000,SchaafTalbot1989, CadilheAraujoPrivman2007}.

The clearest example is R\'enyi's one-dimensional car-parking problem \cite{Renyi1958}. Unit segments are placed sequentially and uniformly on a line, with overlapping attempts rejected. In the infinite-length limit, the jammed density is
\begin{equation}
\rho_{\mathrm R}
=
\int_{0}^{\infty}
\exp\left[
-2\int_{0}^{v}
\frac{1-e^{-u}}{u}\,\dd u
\right]\dd v
=
0.7475979202\ldots .
\label{eq:renyi_constant}
\end{equation}
This number is not obtained by optimizing a packing arrangement. It emerges from the history-dependent fragmentation of the available line: each accepted segment creates smaller gaps, some of which can never be used again. The gap distribution, finite-interval occupation, approach to jamming, and spatial correlations of this process have been studied from several complementary perspectives \cite{Mackenzie1962,DvoretzkyRobbins1964,Widom1966, Pomeau1980,Swendsen1981,HinrichsenFederJossang1986, BaryshnikovGnedin2001}.

Over time, RSA became a useful framework for problems well beyond car parking. Variants of the model have been used to describe colloidal adsorption, protein binding, surface reactions, granular deposition, and the formation of disordered jammed layers
\cite{Feder1980,Evans1993,Talbot2000,CadilheAraujoPrivman2007}. What makes RSA useful in such different settings is precisely the
separation between a simple local acceptance rule and a nontrivial collective outcome. The final configuration remembers how it was built, even though the individual deposition attempts are usually independent.

This picture is also supported by rigorous probabilistic results. For broad classes of finite-range sequential packing processes,
stabilization methods establish a thermodynamic jamming density, extensive occupation fluctuations, and central-limit behavior
\cite{Penrose2001,PenroseYukich2002, SchreiberPenroseYukich2007}. If \(N_L\) denotes the number of adsorbed particles in an interval of length \(L\), one typically finds $\mathbb{E}N_L\sim\rho_\infty L$ and $\operatorname{Var}(N_L)\sim\chi_\infty L$.

The fluctuations of the intensive density \(N_L/L\) then decay as \(L^{-1/2}\). Similar fluctuation questions have been investigated in homogeneous and heterogeneous RSA, as well as in competitive deposition from particle mixtures \cite{LoscarBorziAlbano2003,SubashievLuryi2007}. These results provide a natural reference for the present problem, where the sampling measure itself changes with the state produced by earlier depositions.

A central assumption of conventional continuum RSA is that all geometrically admissible deposition coordinates are sampled uniformly. The existing configuration determines where a new particle can fit, but does not otherwise favor one admissible position over another. This assumption is convenient, although it is not natural for many adsorption processes. Surface heterogeneity, localized binding sites, substrate patterning, roughness, external fields, and particle-substrate interactions can all bias the arrival or attachment position.

Several extensions of RSA introduce such effects through a landscape that exists before adsorption begins. Random-site and heterogeneous substrate models assign different accessibility or attachment probabilities to different regions \cite{Jin1993,Adamczyk1994}. Patterned substrates restrict deposition to prescribed landing zones \cite{AraujoCadilhePrivman2008,PrivmanYan2016, VermaPrivman2018}, while substrate topography changes the available geometry itself \cite{KubalaCiesla2022}. In these examples, the non-uniformity is mainly quenched: it belongs to the substrate and is not regenerated by the deposition process. Random defects provide a related mechanism. They modify the decomposition of the free space and, when sufficiently broad or correlated, can alter conventional finite-size fluctuation scaling \cite{LoscarBorziAlbano2003,PalaciosEtAl2024}.

Interactions can enter in a different way. External fields may bias particle transport toward the substrate \cite{Pagonabarraga1995}, while electrostatic double-layer forces influence the deposition of charged colloids \cite{Oberholzer1997}. Cooperative adsorption models allow nearby particles to enhance or inhibit subsequent events \cite{Evans1993,Talbot2000}. Mixtures introduce another source of history dependence because different particle species compete for the same remaining space \cite{TalbotSchaaf1989,SubashievLuryi2007}. Exact one-dimensional results are available for some interaction rules and even for finite-range potentials designed to optimize the final coverage \cite{Baule2019}. In most of these approaches, however, interactions modify a trial rate, an acceptance probability, or the motion that precedes irreversible attachment.

The mechanism considered here is different. We prescribe the normalized conditional density of the deposition coordinate inside each available gap. This density depends jointly on the charges bounding the gap and on the endpoint charges of the incoming segment. The disorder is therefore not fixed in advance. Every accepted particle creates two daughter gaps with new charge states, and those states determine the deposition measures used in the next generation. In this sense, geometry and charge evolve together as a multitype random fragmentation process \cite{Bertoin2006}.

This distinction matters because the final coverage alone does not describe the process completely. The occupation distribution contains finite-size probabilities, fluctuations, and higher cumulants that cannot generally be reconstructed from its mean. Exact
full-distribution results remain uncommon, particularly for continuum RSA. They are better understood for selected interval-packing and lattice models \cite{BaryshnikovGnedin2001,Krapivsky2020}. A state-dependent continuum kernel adds another layer to the problem: after each deposition, the two daughter gaps inherit different charge states and therefore follow different occupation laws.

We first explored this idea in a model containing only the two dipolar orientations \(+-\) and \(-+\) \cite{PalaciosMacedo2025}. There, the polarization of the parent gap selected a nonuniform placement rule, and the incoming segment relaxed toward an energetically preferred position. Exact recurrences were obtained for the mean occupation, and the variance was separated into contributions generated within and between the deposition channels. That model showed that charge-conditioned RSA can remain analytically tractable after the uniform-sampling assumption is abandoned. Its deterministic placement rules, however, represented a singular limit: once a channel was selected, the deposition coordinate was fixed. The model therefore did not provide a regular interpolation between uniform RSA and strongly localized deposition, nor did it describe the complete occupation distribution.

In the present work, we replace those deterministic rules by a continuous family of normalized beta kernels. Their parameters are set by the charges at the two boundaries of the parent gap and at the two endpoints of the incoming segment. The interaction parameter continuously deforms uniform car parking into increasingly localized and charge-selective deposition, while the boundary of its parameter range recovers deterministic endpoint measures in the corresponding charge sectors. The beta family is flexible enough to describe attractive, suppressive, and asymmetric placement, but still simple enough to preserve an exact recursive decomposition.

We use this structure to derive a four-component probability-generating-function equation for the complete finite-size occupation statistics. From it, we obtain closed recursions for the mean and variance, construct exact finite-shell solutions, and develop
Gauss--Jacobi and Monte Carlo methods for larger systems. We also show that the influence of the initial boundary charges disappears from the extensive mean and variance coefficients, although the interaction strength and particle composition continue to control their values. The resulting model provides a tractable connection between uniform  continuum RSA, dynamically generated disorder, and the singular charged-segment dynamics studied previously.

\section{Model}
In the present work we introduce a continuous family of normalized charge-dependent deposition kernels. Consider an available interval of length \(x\), with boundary charges \(\alpha,\beta\in\{+,-\}\), and an incoming unit segment with ordered endpoint charges \(\sigma,\tau\in\{+,-\}\). If \(u\in(0,x-1)\) denotes the position of the left endpoint of the deposited segment, its conditional density is chosen as
\begin{equation}
P_{\alpha\beta}^{\sigma\tau}(u|x)
=
\frac{
u^{a_{\alpha\sigma}-1}
(x-1-u)^{b_{\tau\beta}-1}
}{
(x-1)^{a_{\alpha\sigma}+b_{\tau\beta}-1}
B(a_{\alpha\sigma},b_{\tau\beta})
},
\label{eq:introduction_beta_kernel}
\end{equation}
where
\begin{equation}
a_{\alpha\sigma}
=
1+\lambda q_\alpha q_\sigma,
\qquad
b_{\tau\beta}
=
1+\lambda q_\tau q_\beta,
\label{eq:introduction_beta_parameters}
\end{equation}
and
\begin{equation}
q_+=1,\quad q_-=-1.
\label{eq:introduction_beta_parameter1s}
\end{equation}
The parameter \(0\leq\lambda<1\) continuously controls the strength of the charge-dependent bias. At \(\lambda=0\), all beta parameters equal one and Eq.~\eqref{eq:introduction_beta_kernel} reduces to the uniform car-parking kernel. For \(\lambda>0\), like-charge contacts are suppressed and opposite-charge contacts are enhanced, while the kernel remains exactly normalized.

\begin{figure}[htbp]
    \centering

    \begin{subfigure}[t]{0.48\linewidth}
        \centering
        \includegraphics[width=\linewidth]{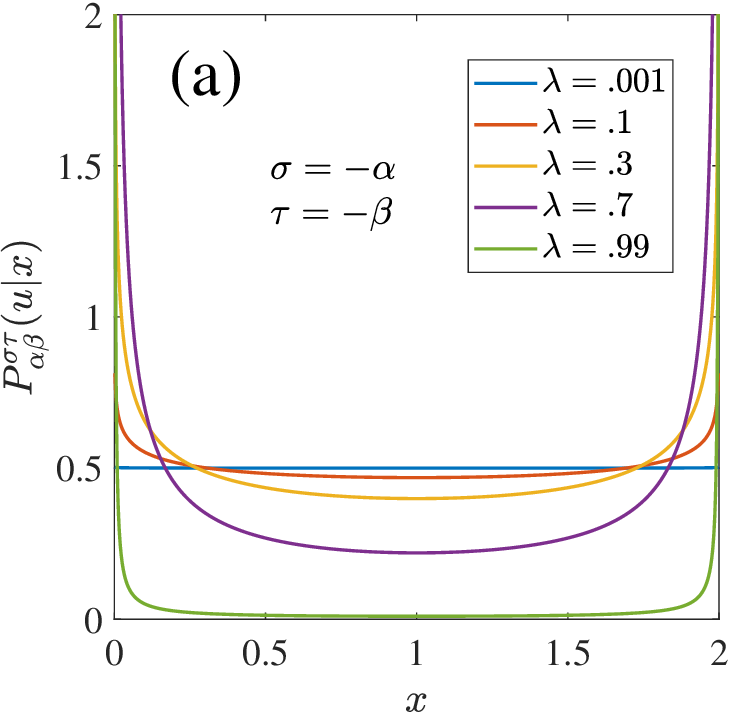}
    \end{subfigure}%
    \hfill
    \begin{subfigure}[t]{0.48\linewidth}
        \centering
        \includegraphics[width=\linewidth]{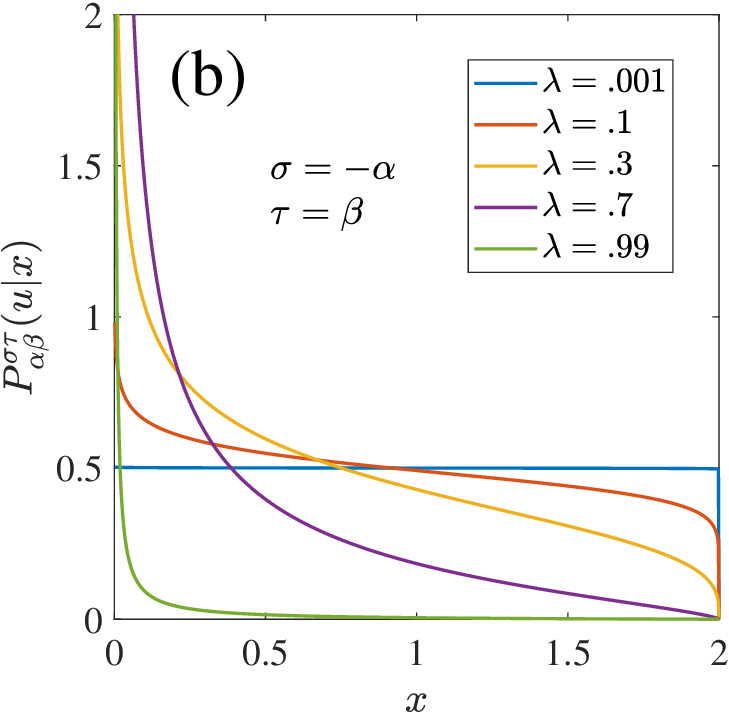}
    \end{subfigure}

    \medskip

    \begin{subfigure}[t]{0.48\linewidth}
        \centering
        \includegraphics[width=\linewidth]{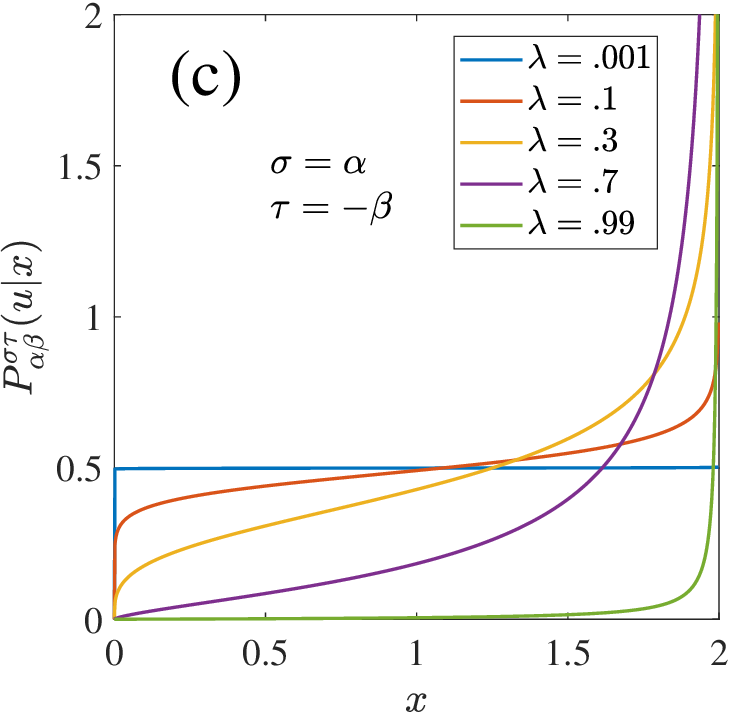}
    \end{subfigure}%
    \hfill
    \begin{subfigure}[t]{0.48\linewidth}
        \centering
        \includegraphics[width=\linewidth]{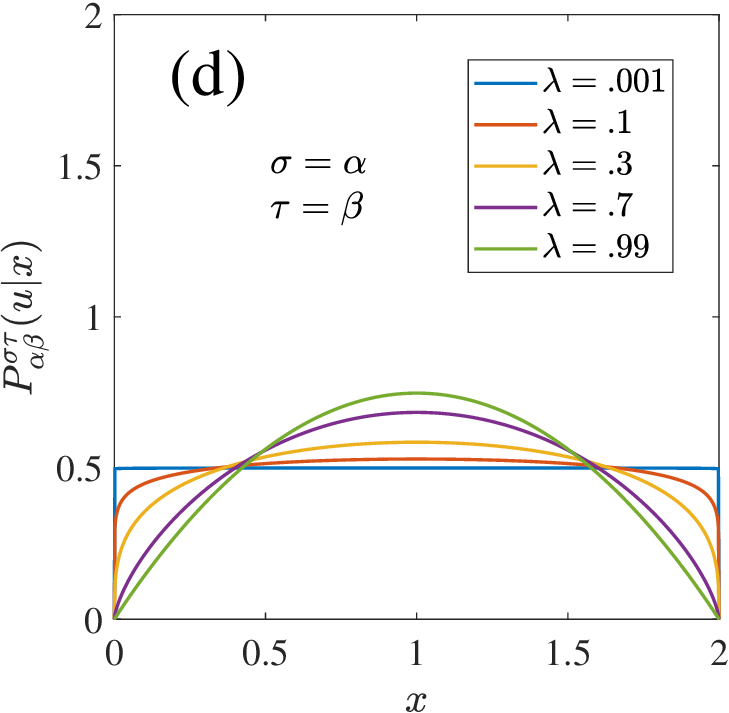}
    \end{subfigure}

    \caption{\protect\justifying\footnotesize Charge-dependent deposition kernels for representative values of \(\lambda\): (a) opposite charges at both contacts; (b) opposite charges at the left contact and equal charges at the right contact; (c) equal charges at the left contact and opposite charges at the right contact; and (d) equal charges at both contacts.}
    \label{fig:beta_kernel_profiles}
\end{figure}

Figure~\ref{fig:beta_kernel_profiles} illustrates the shape of the deposition kernel for the four possible interaction configurations between the charges bounding the parent gap and the endpoint charges of the incoming segment. These configurations distinguish whether the left and right charge pairs are equal or opposite.

Note that opposite charge pairs enhance deposition near the corresponding boundary, whereas equal-charge pairs suppress it. Increasing \(\lambda\) continuously transforms the uniform car-parking kernel into increasingly charge-selective deposition profiles.

The behavior shown in Fig.~\ref{fig:beta_kernel_profiles} suggests a natural thermal interpretation of the parameter \(\lambda\). At \(\lambda=0\), the deposition coordinate is uniformly distributed, whereas increasing \(\lambda\) progressively localizes the deposition near the energetically favored boundaries. To make this connection explicit, \(\lambda\) must be related to the ratio between the interaction and thermal energy scales.

Writing \(L=x-1\) and \(y=u/L\), consider the effective gap-local interaction
\begin{equation}
U_{\alpha\beta}^{\sigma\tau}(y)
=
-\varepsilon
\left[
q_\alpha q_\sigma\ln y
+
q_\tau q_\beta\ln(1-y)
\right],
\label{eq:logarithmic_potential}
\end{equation}
where \(\varepsilon>0\) sets the interaction-energy scale. Its conditional Boltzmann weight is
\begin{align}
\exp\left[
-\frac{U_{\alpha\beta}^{\sigma\tau}(y)}
{k_{\mathrm B}T}
\right]
&=
y^{\lambda q_\alpha q_\sigma}
(1-y)^{\lambda q_\tau q_\beta},
\label{eq:boltzmann_beta_weight}
\\
\lambda
&=
\frac{\varepsilon}{k_{\mathrm B}T}.
\label{eq:lambda_temperature}
\end{align}
After normalization, this weight gives precisely the beta kernel defined in Eq.~\eqref{eq:introduction_beta_kernel}. Thus, \(\lambda\) is an effective inverse-temperature or coupling parameter: it measures the interaction strength relative to the thermal energy.

In the high-temperature or weak-interaction limit, 
\begin{equation}
T\to\infty
\quad\text{or}\quad
\varepsilon\to0
\qquad\Longrightarrow\qquad
\lambda\to0,
\end{equation}
all beta parameters approach one and the uniform car-parking kernel is recovered. Increasing \(\lambda\) enhances the charge-selective bias and drives the deposition toward the boundaries favored by opposite-charge interactions.

This interpretation is effective rather than microscopic. The potential in Eq.~\eqref{eq:logarithmic_potential} describes only the interaction between the incoming endpoints and the charges bounding the parent gap. More distant charges are assumed to be screened by the previously deposited segments. Moreover, the Boltzmann weight governs only the conditional selection of the deposition coordinate. The global adsorption dynamics remains irreversible, does not satisfy detailed balance, and terminates in a jammed state.

The condition \(\lambda<1\) also acquires a direct thermal interpretation. For an attractive opposite-charge contact, the kernel behaves near the corresponding boundary as
\begin{equation}
P(u)\sim u^{-\lambda}.
\end{equation}
Its normalization involves
\begin{equation}
\int_0^1 u^{-\lambda}\,\dd u
=
\frac{1}{1-\lambda},
\end{equation}
which is finite only for \(\lambda<1\). Using
Eq.~\eqref{eq:lambda_temperature}, this condition becomes
\begin{equation}
T>T_{\mathrm c},
\qquad
T_{\mathrm c}
=
\frac{\varepsilon}{k_{\mathrm B}}.
\label{eq:critical_temperature}
\end{equation}
Therefore, \(\lambda\to1^{-}\), or equivalently \(T\to T_{\mathrm c}^{+}\), is an integrability threshold of the continuous deposition measure. It represents an effective logarithmic-collapse transition rather than the ordinary zero-temperature limit.

The endpoint \(\lambda=1\) is not a regular member of the continuous beta-kernel family, but it possesses well-defined weak limits. With
\begin{equation}
r=1+\lambda,
\qquad
t=1-\lambda,
\label{eq:beta_limit_parameters}
\end{equation}
one obtains
\begin{align}
\operatorname{Beta}(r,t)
&\Longrightarrow \delta(y-1),
\\
\operatorname{Beta}(t,r)
&\Longrightarrow \delta(y),
\\
\operatorname{Beta}(t,t)
&\Longrightarrow
\frac{1}{2}\delta(y)
+
\frac{1}{2}\delta(y-1),
\\
\operatorname{Beta}(r,r)
&\Longrightarrow
\operatorname{Beta}(2,2).
\label{eq:weak_beta_limits}
\end{align}
The first three limits describe complete localization at one or both boundaries and recover the corresponding deterministic deposition rules introduced in our previous charged-segment model \cite{PalaciosMacedo2025}. The \((r,r)\) channel remains continuous, so the previous delta-kernel dynamics is recovered as a singular sector, rather than as the complete \(\lambda=1\) limit, of the present theory.

\section{Full-counting-statistics formulation}
\label{sec:fcs}

Let \(N_{\alpha\beta}(x)\) denote the random number of unit segments present at jamming in a gap of length \(x\) bounded by charges \(\alpha,\beta\in\{+,-\}\). If \(x<1\), no segment can be deposited and therefore
\begin{equation}
N_{\alpha\beta}(x)=0,
\qquad 0\leq x<1.
\label{eq:fcs_empty_gap}
\end{equation}

For \(x\geq1\), let the first deposited segment have endpoint charges \(\sigma,\tau\), selected with probability \(p_{\sigma\tau}\), and let its left endpoint be deposited at the random position \(U\), distributed according to the beta kernel \(P_{\alpha\beta}^{\sigma\tau}(u|x)\). The deposition splits the parent gap into the two daughter gaps
\[
(\alpha\sigma,U),
\qquad
(\tau\beta,x-1-U).
\]
Since an adsorbed segment separates the two daughter intervals, their subsequent adsorption processes are conditionally
independent. We assume that the two endpoint charges are selected independently, with \(P(+)=p\) and \(P(-)=1-p\). Hence,
\begin{equation}
p_{++}=p^2,
\qquad
p_{+-}=p_{-+}=p(1-p),
\qquad
p_{--}=(1-p)^2.
\label{eq:segment_probabilities}
\end{equation}
In particular, \(\sum_{\sigma,\tau}p_{\sigma\tau}=1\), and all four particle types occur with positive probability when \(0<p<1\).
Consequently,
\begin{equation}
N_{\alpha\beta}(x)
\overset{d}{=}
1+
N_{\alpha\sigma}^{(L)}(U)
+
N_{\tau\beta}^{(R)}(x-1-U),
\qquad x\geq1,
\label{eq:fcs_distributional_recursion}
\end{equation}
where the superscripts \(L\) and \(R\) denote conditionally independent realizations.

To encode the complete occupation statistics, we introduce the probability-generating function
\begin{equation}
G_{\alpha\beta}(z,x)
=
\mathbb{E}\!\left[
z^{N_{\alpha\beta}(x)}
\right].
\label{eq:fcs_pgf}
\end{equation}
Conditioning Eq.~\eqref{eq:fcs_distributional_recursion} on the particle polarity and deposition position gives
\begin{equation}
\begin{split}
G_{\alpha\beta}(z,x)
={}&
z\sum_{\sigma,\tau=\pm}
p_{\sigma\tau}
\int_{0}^{x-1}
P_{\alpha\beta}^{\sigma\tau}(u|x)
\\
&\times
G_{\alpha\sigma}(z,u)
G_{\tau\beta}(z,x-1-u)\,
\mathrm{d}u ,
\qquad x\geq1,
\end{split}
\label{eq:fcs_integral_equation}
\end{equation}
with
\begin{equation}
G_{\alpha\beta}(z,x)=1,
\qquad 0\leq x<1.
\label{eq:fcs_boundary_condition}
\end{equation}
The normalization of the beta kernels immediately implies
\[
G_{\alpha\beta}(1,x)=1.
\]
Finally, expand the generating function as
\begin{equation}
G_{\alpha\beta}(z,x)
=
\sum_{k=0}^{\lfloor x\rfloor}
\pi_{\alpha\beta,k}(x)z^k,
\;\;
\pi_{\alpha\beta,k}(x)
=
\mathbb{P}\!\left[
N_{\alpha\beta}(x)=k
\right].
\label{eq:fcs_probability_expansion}
\end{equation}
Extracting the coefficient of \(z^k\) from
Eq.~\eqref{eq:fcs_integral_equation} yields, for \(k\geq1\),
\begin{equation}
\begin{split}
\pi_{\alpha\beta,k}(x)
={}&
\sum_{\sigma,\tau=\pm}
p_{\sigma\tau}
\int_{0}^{x-1}
P_{\alpha\beta}^{\sigma\tau}(u|x)
\\
&\times
\sum_{j=0}^{k-1}
\pi_{\alpha\sigma,j}(u)
\pi_{\tau\beta,k-1-j}(x-1-u)\,
\mathrm{d}u .
\end{split}
\label{eq:fcs_probability_recursion}
\end{equation}
The initial coefficient is
\begin{equation}
\pi_{\alpha\beta,0}(x)
=
\begin{cases}
1, & 0\leq x<1,\\
0, & x\geq1.
\end{cases}
\label{eq:fcs_zero_particle_probability}
\end{equation}
Equations~\eqref{eq:fcs_probability_recursion} and \eqref{eq:fcs_zero_particle_probability} determine recursively the complete probability distribution of the jammed occupation number.

\subsection{Mean occupation}
\label{subsec:mean_variance}

The four mean occupation functions are obtained directly from the first derivative of the probability-generating functions:
\begin{equation}
f_{\alpha\beta}(x)
=
\mathbb{E}[N_{\alpha\beta}(x)]
=
\left.
\frac{\partial G_{\alpha\beta}(z,x)}
{\partial z}
\right|_{z=1}.
\label{eq:mean_from_pgf}
\end{equation}
Differentiating Eq.~\eqref{eq:fcs_integral_equation}, using
\(G_{\alpha\beta}(1,x)=1\), gives
\begin{equation}
\begin{split}
f_{\alpha\beta}(x)
={}&
1+
\sum_{\sigma,\tau=\pm}
p_{\sigma\tau}
\int_0^{x-1}
P_{\alpha\beta}^{\sigma\tau}(u|x)
\\
&\times
\left[
f_{\alpha\sigma}(u)
+
f_{\tau\beta}(\bar u)
\right]\mathrm{d}u,
\end{split}
\label{eq:mean_compact}
\end{equation}
where \(\bar u=x-1-u.\)
The constant term in Eq.~\eqref{eq:mean_compact} is equal to one because all beta kernels are normalized and
\(\sum_{\sigma,\tau}p_{\sigma\tau}=1\). 

\subsection{Second factorial moment and variance}
\label{subsubsec:fcs_variance}

The fluctuation equations follow directly from the same full-counting-statistics formulation. We define the second factorial moment as
\begin{equation}
h_{\alpha\beta}(x)
=
\mathbb{E}\!\left[
N_{\alpha\beta}(x)
\bigl(N_{\alpha\beta}(x)-1\bigr)
\right]
=
\left.
\frac{\partial^2 G_{\alpha\beta}(z,x)}
{\partial z^2}
\right|_{z=1}.
\label{eq:second_factorial_moment}
\end{equation}

To differentiate the right-hand side of Eq.~\eqref{eq:fcs_integral_equation}, consider a single deposition
branch and write
\begin{equation}
\mathcal{G}(z)
=
zG_{\alpha\sigma}(z,u)
G_{\tau\beta}(z,\bar u).
\end{equation}
Its second derivative at \(z=1\) is
\begin{align}
\left.
\frac{\partial^2\mathcal{G}(z)}
{\partial z^2}
\right|_{z=1}
={}&
h_{\alpha\sigma}(u)
+
h_{\tau\beta}(\bar u)
\nonumber\\
&+
2f_{\alpha\sigma}(u)
+
2f_{\tau\beta}(\bar u)
\nonumber\\
&+
2f_{\alpha\sigma}(u)
f_{\tau\beta}(\bar u),
\label{eq:branch_second_derivative}
\end{align}
where the normalization
\(G_{\gamma\delta}(1,x)=1\) has been used.

It follows that the second factorial moment for an arbitrary boundary-charge configuration \(\alpha\beta\) satisfies
\begin{equation}
\begin{split}
h_{\alpha\beta}(x)
={}&
\sum_{\sigma,\tau=\pm}
p_{\sigma\tau}
\int_0^{x-1}
P_{\alpha\beta}^{\sigma\tau}(u|x)
\Bigl\{
h_{\alpha\sigma}(u)
+h_{\tau\beta}(\bar u)
\\
&
+2\bigl[
f_{\alpha\sigma}(u)
+f_{\tau\beta}(\bar u)
+f_{\alpha\sigma}(u)
 f_{\tau\beta}(\bar u)
\bigr]
\Bigr\}\mathrm{d}u ,
\end{split}
\label{eq:second_factorial_recursion}
\end{equation}

The ordinary second moment is related to the factorial moment through
\begin{equation}
\mathbb{E}[N_{\alpha\beta}^{2}(x)]
=
h_{\alpha\beta}(x)+f_{\alpha\beta}(x).
\label{eq:ordinary_second_moment}
\end{equation}
Therefore, the variance functions are
\begin{align}
s_{\alpha\beta}(x)
&=
h_{\alpha\beta}(x)
+
f_{\alpha\beta}(x)
-
f_{\alpha\beta}^{2}(x),
\label{eq:fcs_variance_alphabeta}
\end{align}

Since \(N_{\alpha\beta}(x)\) is deterministic for \(x<2\),
\begin{equation}
h_{\alpha\beta}(x)=0,
\qquad
s_{\alpha\beta}(x)=0,
\qquad
0\leq x<2.
\label{eq:factorial_variance_boundary}
\end{equation}

Equations~\eqref{eq:mean_compact} and \eqref{eq:second_factorial_recursion} form a closed recursive system for the mean and the second factorial moment. The complete variance is then obtained algebraically from Eqs.~\eqref{eq:fcs_variance_alphabeta}. For a direct recursive evaluation, it is convenient to introduce
\begin{align}
D_{\alpha\beta}^{\sigma\tau}(u;x)
&=
f_{\alpha\sigma}(u)
+
f_{\tau\beta}(\bar u),
\label{eq:D_definition}
\\
W_{\alpha\beta}^{\sigma\tau}(u;x)
&=
s_{\alpha\sigma}(u)
+
s_{\tau\beta}(\bar u),
\label{eq:W_definition}
\end{align}
together with
\begin{equation}
\begin{split}
M_{\alpha\beta}(x)
&=
\sum_{\sigma,\tau=\pm}
p_{\sigma\tau}
\int_0^{x-1}
P_{\alpha\beta}^{\sigma\tau}(u|x)
D_{\alpha\beta}^{\sigma\tau}(u;x)\,
\mathrm{d}u\\
&=
f_{\alpha\beta}(x)-1.
\end{split}
\label{eq:M_definition}
\end{equation}
Reorganizing the preceding FCS relations then gives the four variance equations in the single compact form
\begin{equation}
\begin{split}
s_{\alpha\beta}(x)
={}&
\sum_{\sigma,\tau=\pm}
p_{\sigma\tau}
\int_0^{x-1}
P_{\alpha\beta}^{\sigma\tau}(u|x)
\\
&\times
\left\{
W_{\alpha\beta}^{\sigma\tau}(u;x)
+
\left[
D_{\alpha\beta}^{\sigma\tau}(u;x)
-
M_{\alpha\beta}(x)
\right]^2
\right\}
\mathrm{d}u .
\end{split}
\label{eq:variance_compact}
\end{equation}

The term \(W_{\alpha\beta}^{\sigma\tau}\) propagates the fluctuations already present in the daughter gaps, while the centered-square term
accounts for the fluctuations generated at the current deposition. 

\section{Exact finite-shell solutions}
\label{sec:finite_shells}

The recursive structure of Eq.~\eqref{eq:fcs_integral_equation} provides exact solutions on successive length shells. If \(x\in[n,n+1)\), both daughter lengths are smaller than \(n\), so the solution on the current shell depends only on functions already known at shorter lengths. This construction is exact at any finite \(x\), although closed expressions become increasingly cumbersome beyond the first few shells.

The first two shells are immediate. No segment can be deposited for \(0\leq x<1\), whereas exactly one segment is present for \(1\leq x<2\). Hence,
\begin{equation}
G_{\alpha\beta}(z,x)
=
\begin{cases}
1, & 0\leq x<1,\\
z, & 1\leq x<2,
\end{cases}
\label{eq:G_first_shells}
\end{equation}
with
\begin{equation}
\begin{array}{c|ccc}
 & f_{\alpha\beta}(x) & h_{\alpha\beta}(x)
 & s_{\alpha\beta}(x)\\ \hline
0\leq x<1 & 0 & 0 & 0\\
1\leq x<2 & 1 & 0 & 0
\end{array}.
\label{eq:moments_first_shells}
\end{equation}

The first nontrivial statistics appears for \(2\leq x<3\). Setting \(L=x-1\), the daughter gaps have lengths \(U\) and \(L-U\), with \(1\leq L<2\). They cannot both accommodate another segment, and therefore
\[
N_{\alpha\beta}(x)\in\{1,2\}.
\]
For a branch with beta parameters \(a\) and \(b\), the final occupation is one when both daughter gaps are shorter than one, that is, when \(L-1<U<1\). Since \(Y=U/L\) follows a \(\operatorname{Beta}(a,b)\) distribution, the probability of this event is
\begin{equation}
A_{a,b}(L)
=
I_{1/L}(a,b)-I_{1-1/L}(a,b),
\label{eq:Aab_incomplete_beta}
\end{equation}
where \(I_y(a,b)=B_y(a,b)/B(a,b)\) is the regularized incomplete beta function. Averaging over the particle types gives
\begin{equation}
A_{\alpha\beta}(x)
=
\sum_{\sigma,\tau=\pm}
p_{\sigma\tau}
A_{a_{\alpha\sigma},b_{\tau\beta}}(x-1).
\label{eq:Aalphabeta_definition}
\end{equation}

Consequently,
\begin{equation}
\mathbb{P}[N_{\alpha\beta}(x)=1]
=
A_{\alpha\beta}(x),
\;\
\mathbb{P}[N_{\alpha\beta}(x)=2]
=
1-A_{\alpha\beta}(x),
\end{equation}
and the complete generating function is
\begin{equation}
G_{\alpha\beta}(z,x)
=
A_{\alpha\beta}(x)z
+
\left[1-A_{\alpha\beta}(x)\right]z^2,
\qquad
2\leq x<3.
\label{eq:G_second_shell}
\end{equation}
Its first two moments are therefore
\begin{align}
f_{\alpha\beta}(x)
&=
2-A_{\alpha\beta}(x),
\label{eq:mean_second_shell}
\\
h_{\alpha\beta}(x)
&=
2\left[1-A_{\alpha\beta}(x)\right],
\\
s_{\alpha\beta}(x)
&=
A_{\alpha\beta}(x)
\left[1-A_{\alpha\beta}(x)\right],
\qquad
2\leq x<3.
\label{eq:variance_second_shell}
\end{align}

At \(\lambda=0\), all deposition kernels are uniform and
\begin{equation}
A_{1,1}(L)
=
\frac{2}{L}-1.
\end{equation}
The charge dependence then disappears, yielding
\begin{align}
f(x)
&=
3-\frac{2}{x-1},
\label{eq:uniform_mean_second_shell}
\\
s(x)
&=
\left(\frac{2}{x-1}-1\right)
\left(2-\frac{2}{x-1}\right),
\qquad
2\leq x<3.
\label{eq:uniform_variance_second_shell}
\end{align}
The mean agrees with the corresponding uniform car-parking result
in ~\cite{burridge2004recursive}.

These formulas provide exact benchmarks for the charge inheritance, the generating-function recursion, and the numerical implementation. For \(x\geq3\), the method of steps remains exact, but the resulting integrals no longer reduce to similarly useful closed expressions; the recursion is therefore continued numerically.

\section{Numerical and Monte Carlo solutions}
\label{sec:numerical_solution}
Beyond the second shell, the recursive construction remains valid, but the resulting integrals no longer reduce to useful closed expressions. We therefore continue the same method of steps numerically. At each length \(x\), the mean and variance depend only on daughter intervals of lengths \(u\) and \(x-1-u\), both strictly smaller than \(x\). Consequently, the solution can be computed sequentially on an increasing length grid, using previously determined values of the daughter functions.

The numerical implementation requires the evaluation of the charge-dependent kernel integrals at every grid point. An ordinary quadrature becomes inefficient when a beta parameter is smaller than one, because the corresponding kernel has an integrable endpoint singularity. This behavior becomes increasingly pronounced as \(\lambda\to1^{-}\). We therefore use Gauss--Jacobi quadrature, which incorporates the beta endpoint factors directly into its weights.

Writing \(L=x-1\), \(v=u/L\),
\(a=a_{\alpha\sigma}\), and \(b=b_{\tau\beta}\), a generic kernel
integral takes the form
\begin{equation}
\begin{split}
&\int_0^{x-1}
P_{\alpha\beta}^{\sigma\tau}(u|x)
F(u,x-1-u)\,\mathrm{d}u
\\
&\qquad =
\frac{1}{B(a,b)}
\int_0^1
v^{a-1}(1-v)^{b-1}
F\!\left(Lv,L(1-v)\right)\mathrm{d}v
\\
&\qquad \simeq
\sum_{j=1}^{Q}
\omega_j^{(a,b)}
F\!\left(
Lv_j^{(a,b)},
L[1-v_j^{(a,b)}]
\right).
\end{split}
\label{eq:gauss_jacobi_beta_rule}
\end{equation}
Here, \(v_j^{(a,b)}\) and \(\omega_j^{(a,b)}\) are the normalized Gauss--Jacobi nodes and weights associated with the beta parameters \((a,b)\). 

For each contact, the corresponding beta parameter $a$ and $b$ takes only two possible values: \(1+\lambda\) for equal charges and \(1-\lambda\) for
opposite charges. Consequently, the sixteen combinations of parent-gap and particle polarities involve only the four distinct parameter pairs
\[
(1+\lambda,1+\lambda),\quad
(1+\lambda,1-\lambda),\quad
(1-\lambda,1+\lambda),\quad
(1-\lambda,1-\lambda).
\]
For fixed \(\lambda\) and \(Q\), the nodes and weights of these four Gauss--Jacobi rules were computed once. Since they are defined in the normalized coordinate \(v\in[0,1]\), they can be reused at every length \(x\), with the physical deposition positions obtained from \(u=(x-1)v\).

\noindent
The length variable was discretized on a uniform grid
\begin{equation}
x_i=i\Delta x,
\qquad
\Delta x=\frac{1}{m},
\label{eq:length_grid}
\end{equation}
ensuring that all integer shell boundaries belong to the grid. Daughter functions at off-grid lengths were obtained by linear interpolation. Because these lengths never exceed \(x_i-1\), all required values had already been computed at earlier steps. The implementation was checked against the exact finite-shell results, and numerical convergence was verified by increasing \(Q\) and decreasing \(\Delta x\).

For comparison, the process was also simulated directly by Monte Carlo sampling. Each realization starts from a single gap \((\alpha\beta,x)\). Whenever an active gap has length \(x'\geq1\), a particle type \((\sigma,\tau)\) is drawn with probability \(p_{\sigma\tau}\), and its normalized deposition position is sampled
as
\begin{equation}
Y\sim
\operatorname{Beta}
\left(
a_{\alpha\sigma},
b_{\tau\beta}
\right).
\end{equation}
The physical deposition coordinate is then
\begin{equation}
U=(x'-1)Y,
\end{equation}
and the parent gap is replaced by the two daughter gaps
\begin{equation}
(\alpha\beta,x')
\longrightarrow
(\alpha\sigma,U)
\oplus
(\tau\beta,x'-1-U).
\end{equation}
The procedure is repeated recursively until all remaining gaps have length smaller than one. The total number of accepted segments gives
one realization of \(N_{\alpha\beta}(x)\), and independent realizations provide its probability distribution, mean, and variance.

Because the deposition position is sampled directly from the normalized conditional kernel, no rejected trial positions are required. The Monte Carlo simulation therefore follows exactly the same recursive dynamics as the generating-function formulation and provides an independent check of the quadrature results.

This rejection-free construction should not be interpreted as a time-resolved simulation of deposition attempts over the complete substrate. In such a kinetic description, unsuccessful trials would determine the waiting times and would become increasingly frequent as the system approaches jamming. Here, however, the kernel is defined as the conditional distribution of an accepted deposition inside an available gap, and the objective is only to sample the final jammed state. Rejected attempts can therefore be integrated out without changing the sequence of accepted configurations or the resulting occupation statistics.

\begin{figure}[h]
    \centering

    \begin{subfigure}[t]{0.48\linewidth}
        \centering
        \includegraphics[width=\linewidth]{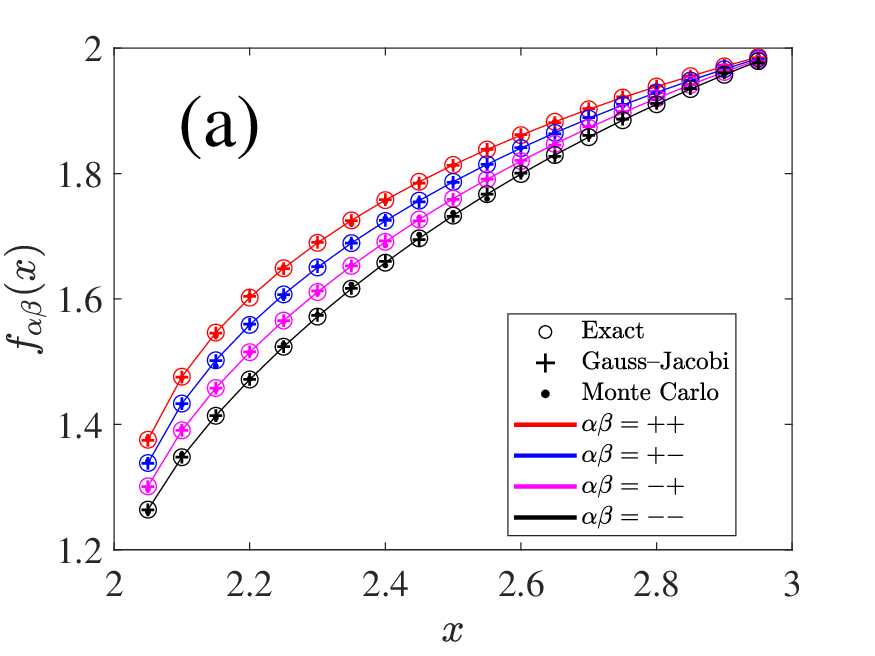}
    \end{subfigure}%
    \hfill
    \begin{subfigure}[t]{0.48\linewidth}
        \centering
        \includegraphics[width=\linewidth]{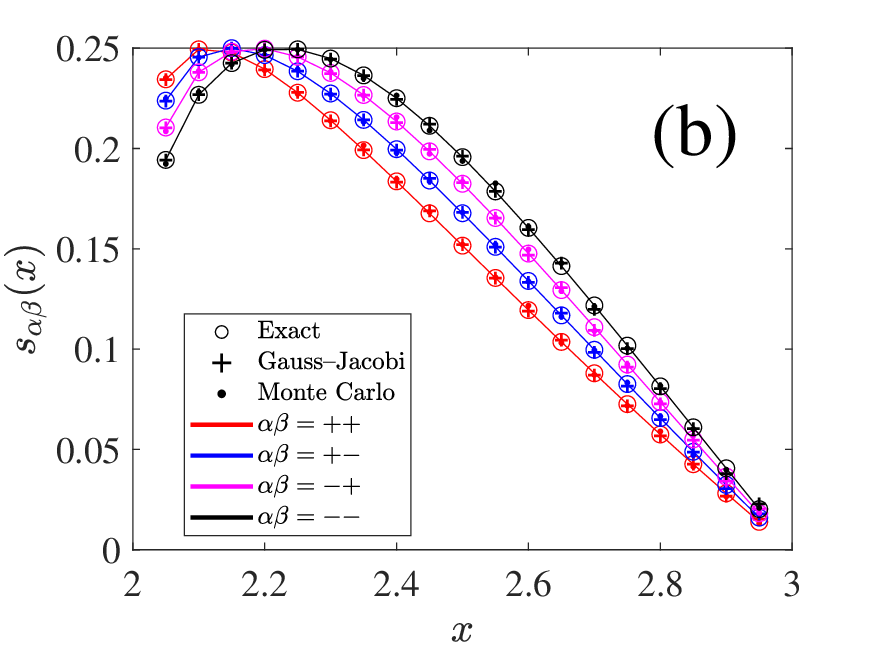}
    \end{subfigure}

    \medskip

    \begin{subfigure}[t]{0.48\linewidth}
        \centering
        \includegraphics[width=\linewidth]{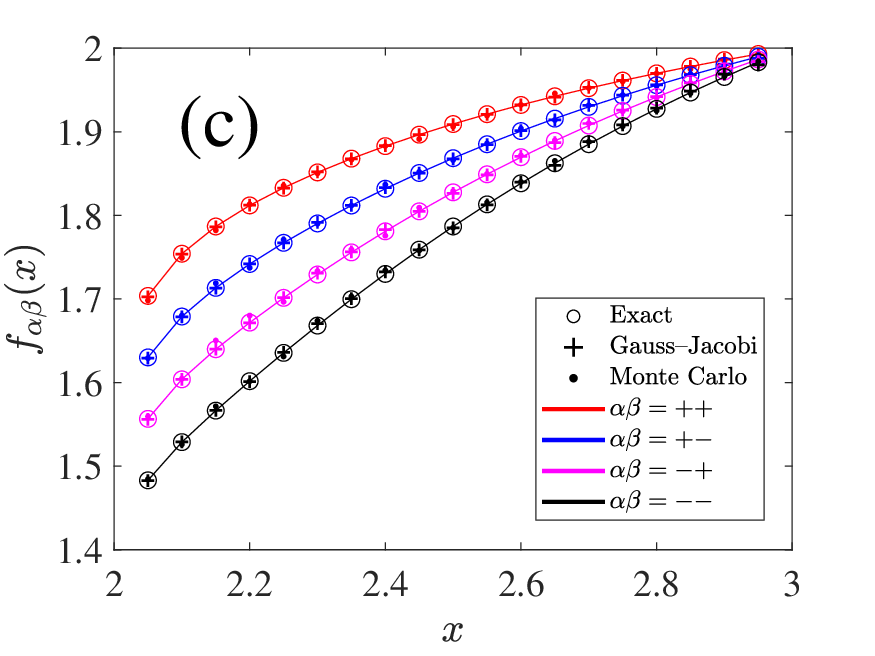}
    \end{subfigure}%
    \hfill
    \begin{subfigure}[t]{0.48\linewidth}
        \centering
        \includegraphics[width=\linewidth]{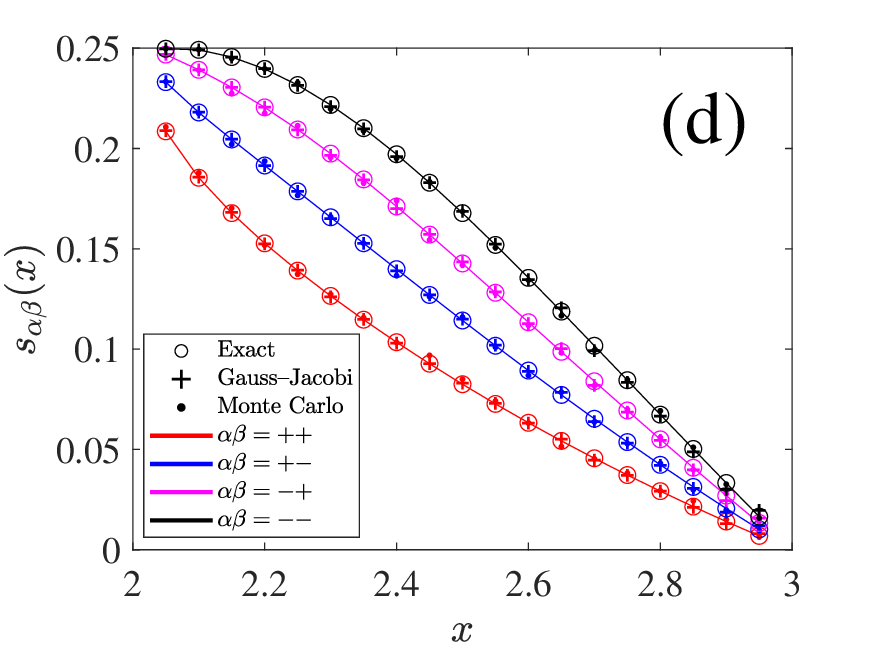}
    \end{subfigure}

    \caption{\protect\justifying\footnotesize Mean occupation \(f_{\alpha\beta}(x)\) and variance \(s_{\alpha\beta}(x)\) in the first nontrivial shell \(2\leq x<3\), for \(\lambda=0.65\) (top row) and \(\lambda=0.90\) (bottom row). The four colors represent the boundary polarities \(\alpha\beta\), while circles, crosses, and dots denote, respectively, the exact method-of-steps solution, Gauss--Jacobi quadrature, and Monte Carlo simulations.}
    \label{fig:method_comparison}
\end{figure}

Figure~\ref{fig:method_comparison} compares the exact method-of-steps solution, Gauss--Jacobi quadrature, and Monte Carlo simulations throughout the second shell, \(2\leq x<3\), for the four boundary polarities and for \(\lambda=0.65\) and \(0.90\).

\noindent 
The calculations shown in Fig.~\ref{fig:method_comparison} were performed using
\[
\textbf{p}=(p_{++},p_{+-},p_{-+},p_{--})=(0.1,0.2,0.3,0.4),
\]
with \(\Delta x=0.05\), \(Q=2048\) Gauss--Jacobi nodes, and \(N_{\mathrm{rep}}=10^{7}\) independent Monte Carlo realizations.

The physical effect of increasing \(\lambda\) is clearly visible when the two scenarios of values of $\lambda$ are compared. Because negatively charged segments are the most probable particles, \(p_{--}=0.4\), a \(++\) parent gap favors opposite-charge contacts and therefore deposition near its boundaries. This produces a larger mean occupation and smaller fluctuations. The opposite behavior occurs for a \(--\) parent gap, where equal-charge contacts suppress boundary deposition and shift the kernel toward the gap interior. These differences are moderate at \(\lambda=0.65\) but become considerably stronger at \(\lambda=0.90\), as the deposition rule becomes more charge selective. Nevertheless, as \(x\to3^{-}\), all four means approach \(2\) and the corresponding variances vanish, as required by the second-shell geometry.

Using the same parameters, Table~\ref{tab:exact_GJ_MC_comparison}
reports the corresponding mean and variance at the representative
length \(x=2.5\), separately for each boundary polarity and both
values of \(\lambda\).

\begin{table}[h]
\centering
\caption{Comparison of the exact, Gauss--Jacobi, and Monte Carlo results at $x=2.50$.}
\label{tab:exact_GJ_MC_comparison}
\begin{tabular}{c c ccc ccc}
\toprule
$\lambda$ & $\alpha\beta$ & $f_{\mathrm{exact}}$ & $f_{\mathrm{GJ}}$ & $f_{\mathrm{MC}}$ & $s_{\mathrm{exact}}$ & $s_{\mathrm{GJ}}$ & $s_{\mathrm{MC}}$ \\
\midrule
0.65 & $++$ & 1.813714 & 1.81356 & 1.81359 & 0.15158 & 0.15168 & 0.15166 \\
 & $+-$ & 1.786823 & 1.78665 & 1.78667 & 0.16773 & 0.16783 & 0.16782 \\
 & $-+$ & 1.759931 & 1.75973 & 1.76014 & 0.18244 & 0.18254 & 0.18233 \\
 & $--$ & 1.733040 & 1.73282 & 1.73295 & 0.19569 & 0.19580 & 0.19574 \\
\midrule
0.90 & $++$ & 1.909274 & 1.90928 & 1.90933 & 0.08249 & 0.08249 & 0.08245 \\
 & $+-$ & 1.868418 & 1.86835 & 1.86838 & 0.11427 & 0.11431 & 0.11430 \\
 & $-+$ & 1.827562 & 1.82743 & 1.82759 & 0.14270 & 0.14279 & 0.14269 \\
 & $--$ & 1.786705 & 1.78650 & 1.78680 & 0.16780 & 0.16792 & 0.16775 \\
\bottomrule
\end{tabular}
\end{table}

Both numerical methods reproduce the exact mean and variance within a few \(10^{-4}\) in every case. The agreement remains equally good for \(\lambda=0.90\), despite the stronger concentration of the beta kernels near the gap boundaries. Increasing \(\lambda\) raises the mean occupation and reduces its variance for all four boundary polarities, while preserving their distinct ordering. These results validate the charge-dependent recursion, the Gauss--Jacobi implementation, and the Monte Carlo algorithm.

\section{Asymptotic behavior}
\label{sec:asymptotic_fluctuations}

We now consider the limit \(x\to\infty\) at fixed \(\mathbf{p}\) and
\(\lambda\). Since the deposition interaction is local, we assume the
standard extensive behavior
\begin{equation}
f_{\alpha\beta}(x)
=
\rho_{\alpha\beta}x+O(1),
\qquad
s_{\alpha\beta}(x)
=
\chi_{\alpha\beta}x+O(1),
\label{eq:asymptotic_mean_variance}
\end{equation}
where \(\rho_{\alpha\beta}\) and \(\chi_{\alpha\beta}\) are,
respectively, the asymptotic mean and variance densities. It follows
that
\begin{equation}
\operatorname{Var}\left[
\frac{N_{\alpha\beta}(x)}{x}
\right]
=
\frac{s_{\alpha\beta}(x)}{x^{2}}
=
\frac{\chi_{\alpha\beta}}{x}
+O(x^{-2}).
\label{eq:density_variance_scaling}
\end{equation}
Thus, the absolute fluctuations of \(N_{\alpha\beta}(x)\) grow as
\(\sqrt{x}\), whereas the fluctuations of the occupation density decay
as \(x^{-1/2}\).

Moreover, since
\(f_{\alpha\beta}(x)/x=\rho_{\alpha\beta}+O(x^{-1})\), Chebyshev's
inequality gives, for every \(\varepsilon>0\) and sufficiently large
\(x\),
\begin{equation}
\begin{split}
\mathbb{P}\left(
\left|
\frac{N_{\alpha\beta}(x)}{x}
-\rho_{\alpha\beta}
\right|>\varepsilon
\right)
&\leq
\frac{4s_{\alpha\beta}(x)}{\varepsilon^{2}x^{2}}
\\
&=O(x^{-1})
\longrightarrow0.
\end{split}
\label{eq:chebyshev_density}
\end{equation}
Therefore, the jammed occupation density is self-averaging:
\begin{equation}
\frac{N_{\alpha\beta}(x)}{x}
\xrightarrow[x\to\infty]{\mathbb{P}}
\rho_{\alpha\beta}.
\label{eq:convergence_probability}
\end{equation}
This conclusion requires only the extensive scaling in
Eq.~\eqref{eq:asymptotic_mean_variance}; it neither assumes a central
limit theorem nor implies an asymptotically Gaussian distribution.

A more general derivation can be developed directly from the
generating functions \(G_{\alpha\beta}(z,x)\), showing that the
asymptotic densities of all cumulants are independent of the boundary
charges. Here we restrict the explicit argument to the first two
cumulants, the mean and the variance.

Assuming that the limits
\begin{equation}
\rho_{\alpha\beta}
=
\lim_{x\to\infty}\frac{f_{\alpha\beta}(x)}{x}
\label{eq:def_asymptotic_density}
\end{equation}
exist, divide Eq.~\eqref{eq:mean_compact} by \(x\), set
\(L=x-1\) and \(y=u/L\), and take \(x\to\infty\). Since
\begin{equation}
\frac{f_{\alpha\sigma}(Ly)}{x}
\longrightarrow y\rho_{\alpha\sigma},
\qquad
\frac{f_{\tau\beta}(L[1-y])}{x}
\longrightarrow(1-y)\rho_{\tau\beta},
\end{equation}
we obtain
\begin{equation}
\rho_{\alpha\beta}
=
\sum_{\sigma,\tau=\pm}p_{\sigma\tau}
\left[
\mu_{\alpha\beta}^{\sigma\tau}\rho_{\alpha\sigma}
+
\left(1-\mu_{\alpha\beta}^{\sigma\tau}\right)
\rho_{\tau\beta}
\right],
\label{eq:density_fixed_point}
\end{equation}
where the first moment of the normalized beta kernel is
\begin{equation}
\mu_{\alpha\beta}^{\sigma\tau}
=
\mathbb E[Y]
=
\frac{a_{\alpha\sigma}}
{a_{\alpha\sigma}+b_{\tau\beta}}
=
\frac{1+\lambda q_\alpha q_\sigma}
{2+\lambda(q_\alpha q_\sigma+q_\tau q_\beta)}.
\label{eq:kernel_first_moment}
\end{equation}
Let
\begin{equation}
\boldsymbol{\rho}
=
(\rho_{++},\rho_{+-},\rho_{-+},\rho_{--})^{\mathsf T}.
\end{equation}
Equations~\eqref{eq:density_fixed_point} can be written as
\begin{equation}
\boldsymbol{\rho}=\mathsf{T}\boldsymbol{\rho},
\label{eq:density_matrix_equation}
\end{equation}
with
\begin{equation}
\mathsf{T}_{(\alpha\beta),(\gamma\delta)}
=
\sum_{\sigma,\tau=\pm}p_{\sigma\tau}
\left[
\mu_{\alpha\beta}^{\sigma\tau}
\delta_{\gamma\alpha}\delta_{\delta\sigma}
+
\left(1-\mu_{\alpha\beta}^{\sigma\tau}\right)
\delta_{\gamma\tau}\delta_{\delta\beta}
\right].
\label{eq:transition_matrix}
\end{equation}
Every row of \(\mathsf{T}\) sums to one because
\begin{equation}
\sum_{\gamma,\delta}
\mathsf{T}_{(\alpha\beta),(\gamma\delta)}
=
\sum_{\sigma,\tau}p_{\sigma\tau}
\left[
\mu_{\alpha\beta}^{\sigma\tau}
+1-\mu_{\alpha\beta}^{\sigma\tau}
\right]
=1.
\end{equation}
Thus, \(\mathsf{T}\) is row-stochastic and
\begin{equation}
\mathsf{T}(1,1,1,1)^{\mathsf T}
=(1,1,1,1)^{\mathsf T}.
\end{equation}

For \(0<p<1\), all probabilities in
Eq.~\eqref{eq:segment_probabilities} are positive. Moreover,
\(0\leq\lambda<1\) implies
\begin{equation}
0<\mu_{\alpha\beta}^{\sigma\tau}<1.
\end{equation}
Any boundary state \((\alpha,\beta)\) can therefore reach any other
state \((\gamma,\delta)\) in two steps,
\begin{equation}
(\alpha,\beta)
\longrightarrow(\alpha,\delta)
\longrightarrow(\gamma,\delta),
\end{equation}
with strictly positive weight. Hence, every element of \(\mathsf{T}^{2}\) is positive and \(\mathsf{T}\) is primitive. By the Perron--Frobenius theorem \cite{Seneta1981}, its eigenvalue one is simple. The fixed vector in Eq.~\eqref{eq:density_matrix_equation} must consequently be proportional to the constant vector, which proves that
\begin{equation}
\rho_{++}=\rho_{+-}=\rho_{-+}=\rho_{--}\equiv\rho.
\label{eq:four_equal_densities}
\end{equation}
Equivalently,
\begin{equation}
\lim_{x\to\infty}\frac{f_{\alpha\beta}(x)}{x}
=\rho,
\qquad
\forall\alpha,\beta\in\{+,-\}.
\label{eq:four_equal_limits}
\end{equation}

We now assume that the variance densities 
\begin{equation}
\chi_{\alpha\beta}
=
\lim_{x\to\infty}\frac{s_{\alpha\beta}(x)}{x}
\label{eq:def_variance_density}
\end{equation}
exist. Conditioned on the first particle type and its deposition position, the daughter gaps evolve independently. The law of total variance gives
\begin{align}
s_{\alpha\beta}(x)
={}&
\mathbb E\!\left[
\operatorname{Var}
\left(N_{\alpha\beta}(x)\mid\sigma,\tau,U\right)
\right]
\nonumber\\
&+
\operatorname{Var}\!\left[
\mathbb E
\left(N_{\alpha\beta}(x)\mid\sigma,\tau,U\right)
\right].
\label{eq:total_variance_asymptotic}
\end{align}
The first term propagates the daughter-gap variances. Since the four mean densities are equal, the extensive part of the conditional mean is
\begin{equation}
\rho U+\rho(x-1-U)=\rho(x-1),
\end{equation}
which is independent of \(U\), \(\sigma\), and \(\tau\). The second term in Eq.~\eqref{eq:total_variance_asymptotic} therefore contains no
contribution of order \(x^2\). Retaining the terms linear in \(x\) gives
\begin{equation}
\chi_{\alpha\beta}
=
\sum_{\sigma,\tau=\pm}p_{\sigma\tau}
\left[
\mu_{\alpha\beta}^{\sigma\tau}\chi_{\alpha\sigma}
+
\left(1-\mu_{\alpha\beta}^{\sigma\tau}\right)
\chi_{\tau\beta}
\right].
\label{eq:variance_density_system}
\end{equation}
Thus, with
\begin{equation}
\boldsymbol{\chi}
=(\chi_{++},\chi_{+-},\chi_{-+},\chi_{--})^{\mathsf T},
\end{equation}
we have \(\boldsymbol{\chi}=\mathsf{T}\boldsymbol{\chi}\). The
simplicity of the eigenvalue one then yields
\begin{equation}
\chi_{++}=\chi_{+-}=\chi_{-+}=\chi_{--}\equiv\chi.
\label{eq:four_equal_variance_densities}
\end{equation}
Therefore, for every boundary state,
\begin{equation}
f_{\alpha\beta}(x)=\rho x+O(1),
\qquad
s_{\alpha\beta}(x)=\chi x+O(1).
\label{eq:common_asymptotic_mean_variance}
\end{equation}
The boundary charges may affect subleading finite-size corrections, but not the extensive mean and variance coefficients. This result also has a simple interpretation in terms of loss of boundary memory. Each accepted segment screens the parent gap and replaces it by two daughter gaps whose inner boundary charges are determined by the deposited
particle. Repeated deposition therefore progressively replaces the information carried by the initial charges \((\alpha,\beta)\) with charges drawn from the particle distribution \(\mathbf{p}\). For a primitive charge-transition matrix, this mixing erases the dependence on the initial boundary state in the bulk, while any remaining memory is confined to subextensive boundary corrections. This provides a direct physical counterpart to the Perron--Frobenius argument: the macroscopic occupation and its leading fluctuations become independent of \((\alpha,\beta)\).

\begin{figure}[h]
    \centering

    \begin{subfigure}[t]{1\linewidth}
        \centering
        \includegraphics[width=\linewidth]{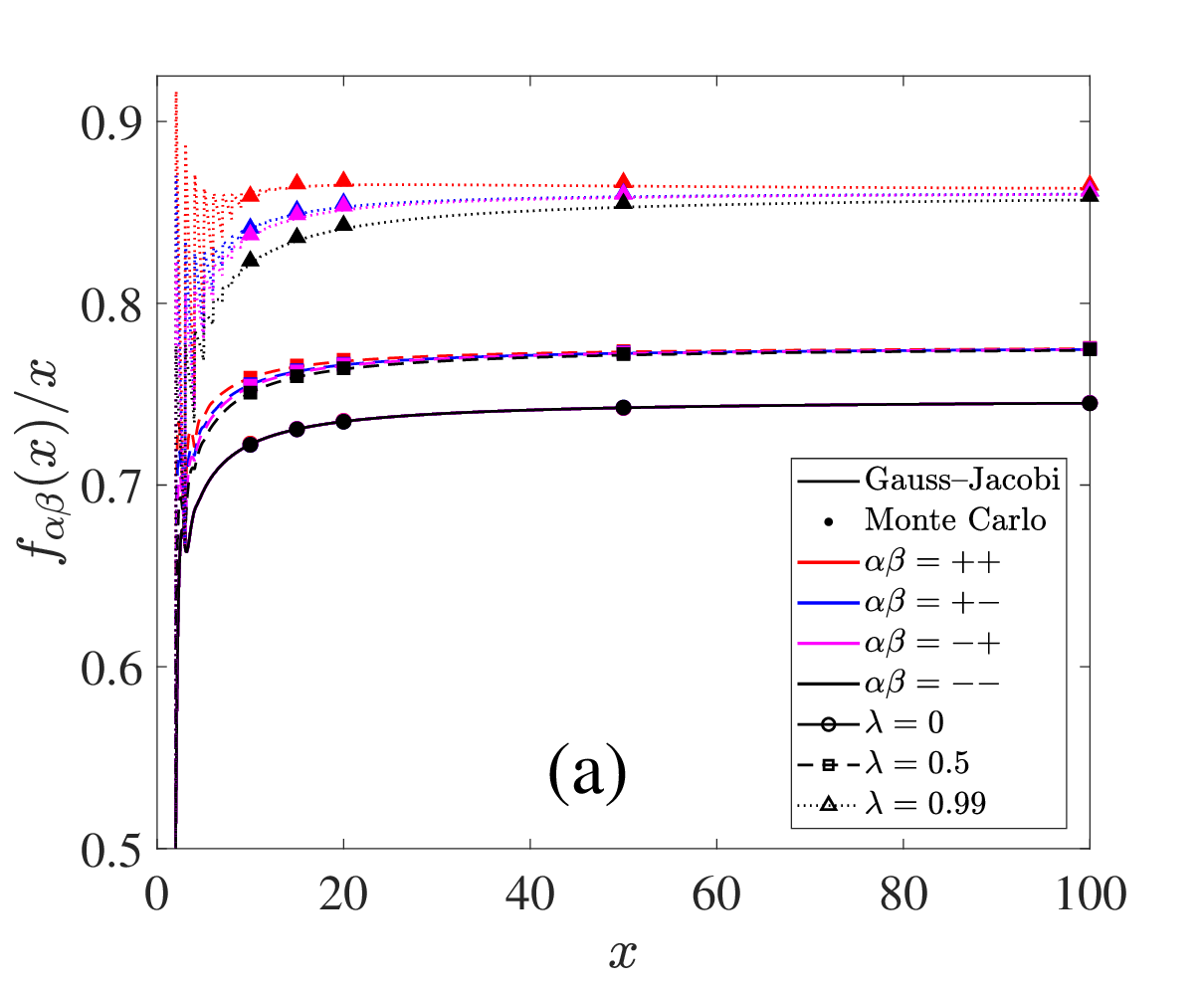}
    \end{subfigure}\\
    \begin{subfigure}[t]{1\linewidth}
        \centering
        \includegraphics[width=\linewidth]{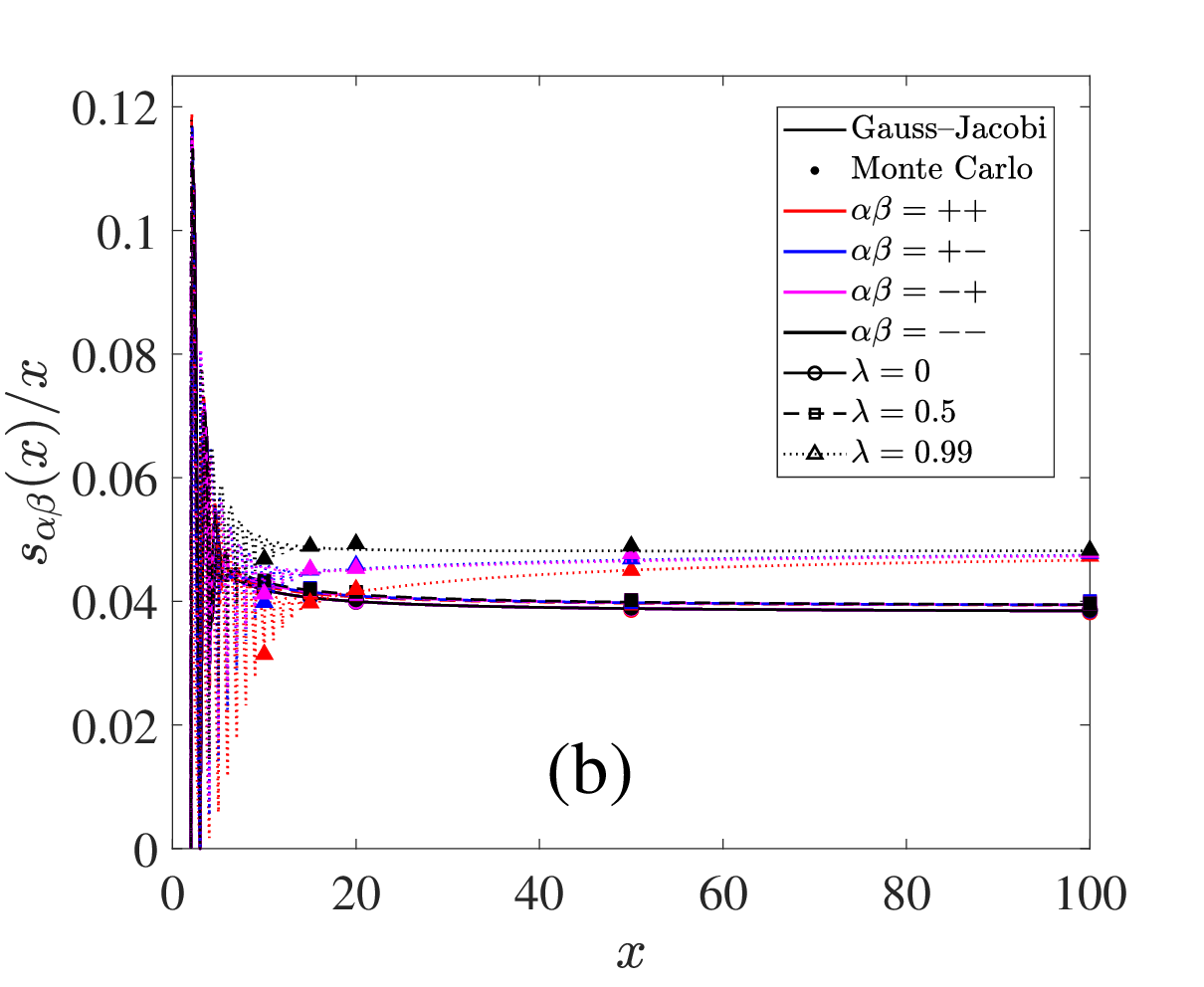}
    \end{subfigure}

    \caption{\protect\justifying\footnotesize Mean occupation density (a) and Variance density (b) for the four boundary polarities and three values of \(\lambda\). Lines represent the Gauss--Jacobi solution and symbols represent Monte Carlo simulations; colors identify the boundary polarity, while line and marker styles identify \(\lambda\).}
    \label{fig:large_length_validation}
\end{figure}

To test the numerical solution beyond the analytically accessible second shell and examine the onset of the asymptotic regime, in Fig. \ref{fig:large_length_validation} we compare Gauss--Jacobi quadrature with independent Monte Carlo simulations up to \(x=100\). The calculations use \(\mathbf{p}=(0.1,0.2,0.3,0.4)\) and three interaction strengths, \(\lambda=0\), \(0.5\), and \(0.99\), ranging from uniform deposition to the vicinity of the singular charge-selective limit.

The Gauss--Jacobi and Monte Carlo results remain in close agreement over the complete length range, including \(\lambda=0.99\), where the beta kernels are strongly concentrated near the gap boundaries. At \(\lambda=0\), the kernel is uniform and the four boundary polarities
are indistinguishable. Charge-dependent differences emerge for \(\lambda>0\), but progressively decrease with \(x\), and all four curves approach a common plateau for each fixed \(\lambda\). This provides a direct numerical manifestation of the loss of boundary memory established by the asymptotic analysis. For the particle distribution considered here, increasing \(\lambda\) raises both the asymptotic occupation density and the variance coefficient. The latter does not contradict self-averaging: although \(s_{\alpha\beta}(x)/x\to\chi\), the fluctuations of the intensive occupation \(N_{\alpha\beta}(x)/x\) still decay as \(x^{-1/2}\)

Having established that the initial boundary charges do not affect the extensive coefficients, we now examine how the asymptotic densities depend on the interaction strength and on the composition of the deposited particles. This distinction is physically important: \(\lambda\) controls the strength of the charge-selective deposition, whereas \(\mathbf{p}\) determines how frequently each type of local charge environment is generated during the recursive fragmentation of the gaps.

We consider three representative particle distributions,
\begin{equation}
\begin{aligned}
\mathbf{p}_{U}&=(0.25,0.25,0.25,0.25),\\
\mathbf{p}_{E}&=(0.45,0.05,0.05,0.45),\\
\mathbf{p}_{O}&=(0.05,0.45,0.45,0.05),
\end{aligned}
\label{eq:pvalues}
\end{equation}
corresponding, respectively, to a uniform mixture, a mixture dominated by equal endpoint charges, and a mixture dominated by opposite endpoint charges. All probabilities remain nonzero, ensuring that the four charge states remain dynamically connected. These three cases separate the effect of interaction strength from that of particle polarity and reveal how the microscopic charge composition controls the bulk occupation and its fluctuations. Table~\ref{tab:asymptotic_densities} reports the resulting asymptotic mean and variance densities, \(\rho\) and \(\chi\), obtained by extrapolating \(f_{\alpha\beta}(x)/x\) and \(s_{\alpha\beta}(x)/x\) linearly in \(1/x\).

\begin{table}[h]
\centering
\caption{Asymptotic mean and variance densities for $\mathbf{p}$ in (\ref{eq:pvalues}).}
\label{tab:asymptotic_densities}
\begin{tabular}{c cc cc cc}
\toprule
 & \multicolumn{2}{c}{Uniform} & \multicolumn{2}{c}{Equal-charge} & \multicolumn{2}{c}{Opposite-charge} \\
\cmidrule(lr){2-3} \cmidrule(lr){4-5} \cmidrule(lr){6-7}
$\lambda$ & $\rho$ & $\chi$ & $\rho$ & $\chi$ & $\rho$ & $\chi$ \\
\midrule
0.00 & 0.7476 & 0.0381 & 0.7476 & 0.0381 & 0.7476 & 0.0381 \\
0.25 & 0.7514 & 0.0381 & 0.7513 & 0.0381 & 0.7516 & 0.0381 \\
0.50 & 0.7649 & 0.0384 & 0.7637 & 0.0385 & 0.7663 & 0.0384 \\
0.75 & 0.7972 & 0.0392 & 0.7918 & 0.0399 & 0.8044 & 0.0392 \\
0.90 & 0.8401 & 0.0396 & 0.8273 & 0.0429 & 0.8609 & 0.0381 \\
0.99 & 0.8912 & 0.0392 & 0.8670 & 0.0493 & 0.9451 & 0.0243 \\
\bottomrule
\end{tabular}
\end{table}

At \(\lambda=0\), all three particle distributions give \(\rho=0.7476\) and \(\chi=0.0381\). This coincidence is expected
because the deposition kernel is uniform and therefore insensitive to the endpoint charges. The resulting value of \(\rho\) recovers the standard one-dimensional car-parking density. Charge composition becomes relevant only when \(\lambda>0\), and its influence grows strongly as the singular limit \(\lambda\to1^{-}\) is approached.

The mean density increases with \(\lambda\) in all three cases, showing that charge-selective deposition produces geometrically more efficient jammed configurations than uniform placement. Boundary-favored events tend to place an incoming segment close to one side of a gap, leaving a larger daughter interval available for subsequent deposition. By contrast, deposition toward the gap interior more frequently divides the available length into two unusable fragments. This effect is strongest for the opposite-charge-dominated mixture, for which \(\rho\) reaches \(0.9451\) at \(\lambda=0.99\), compared with \(0.8912\) for the uniform mixture and \(0.8670\) for the equal-charge-dominated case.

The variance density reveals a further distinction that is not evident from the mean alone. For the uniform mixture, \(\chi\) remains close to \(0.039\) throughout the investigated range. Equal-charge dominance instead increases \(\chi\), reaching \(0.0493\) at \(\lambda=0.99\), indicating stronger sample-to-sample variability generated by the competition between boundary-suppressed and boundary-favored channels. In the opposite-charge-dominated case, the high-density regime is accompanied by a pronounced reduction of \(\chi\), down to \(0.0243\). Thus, the approach to nearly complete packing becomes simultaneously more efficient and more reproducible, consistently with the concentration of the deposition measure near geometrically favorable positions.

\begin{figure}[h]
    \centering
    \begin{subfigure}[t]{.49\linewidth}
        \centering
        \includegraphics[width=\linewidth]{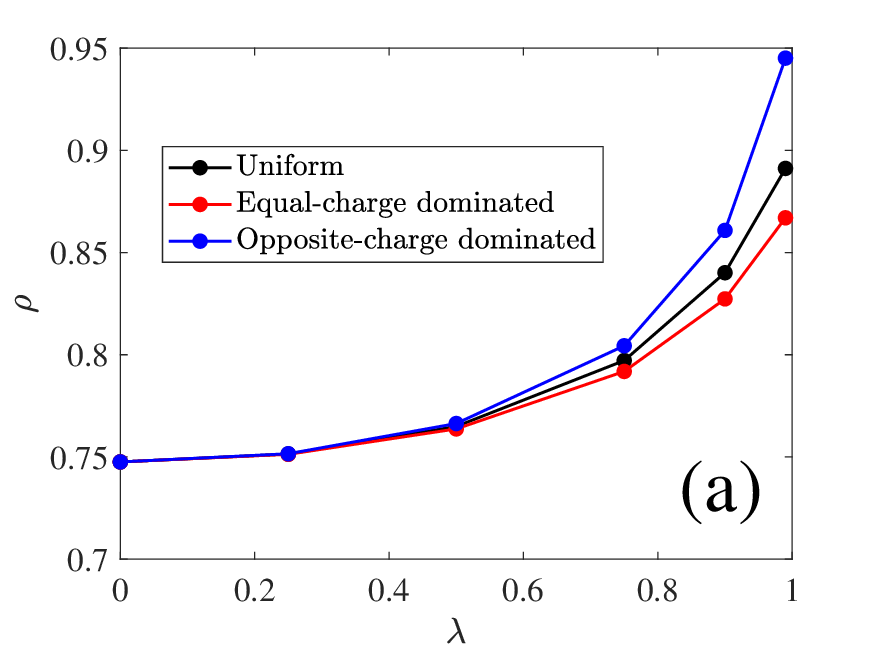}
    \end{subfigure}
    \begin{subfigure}[t]{.49\linewidth}
        \centering
        \includegraphics[width=\linewidth]{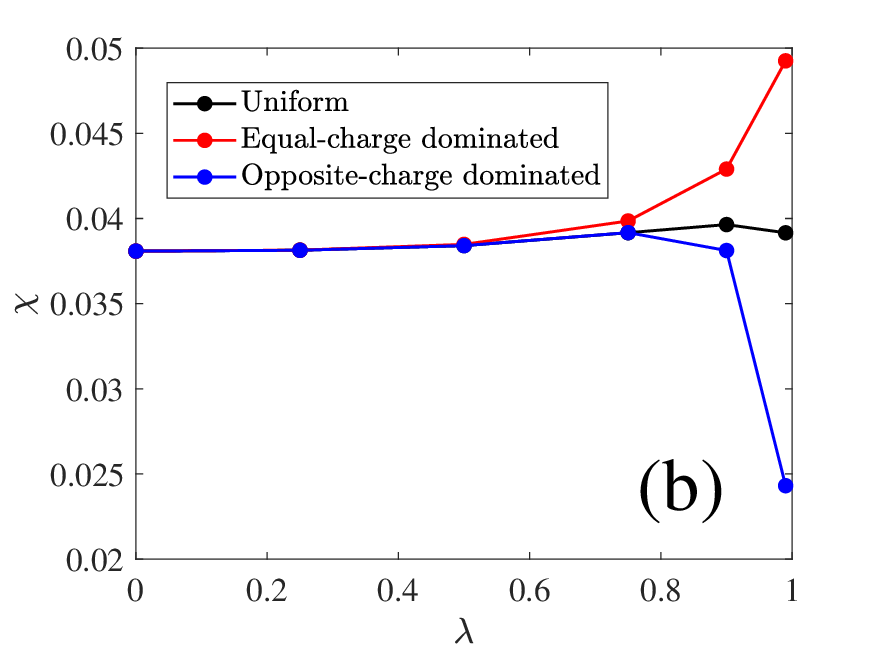}
    \end{subfigure}

    \caption{\protect\justifying\footnotesize Mean occupation density (a) and Variance density (b) for the four boundary polarities and three values of \(\lambda\). Lines represent the Gauss--Jacobi solution and symbols represent Monte Carlo simulations; colors identify the boundary polarity, while line and marker styles identify \(\lambda\).}
    \label{fig:asymptotic_densities}
\end{figure}

The values reported in Table~\ref{tab:asymptotic_densities} are shown in Fig.~\ref{fig:asymptotic_densities} to facilitate their graphical comparison. The figure makes particularly clear that the dependence on particle composition is weak at small \(\lambda\) and becomes strongly nonlinear near \(\lambda=1\). It also highlights the opposite trends of \(\chi\) for equal- and opposite-charge-dominated mixtures, which are less apparent from the individual numerical entries in the table.

\section{Conclusions}
\label{sec:conclusions}

We have developed a generalized one-dimensional RSA model in which the deposition position depends on both the charges bounding the available gap and the endpoint charges of the incoming segment. The beta kernel preserves the recursive structure of the car-parking problem while providing a continuous connection between uniform deposition at \(\lambda=0\) and the strongly localized, charge-selective rules considered in our previous work as \(\lambda\to1^{-}\)~\cite{PalaciosMacedo2025}.

The full counting statistics formulation yields exact recursions for the complete occupation statistics and closed solutions on the first three length shells. For larger intervals, Gauss--Jacobi quadrature accurately handles the endpoint singularities of the kernels, with Monte Carlo simulations providing an independent validation. The asymptotic analysis shows that the initial boundary charges affect only finite-size corrections: all four boundary polarities share the same mean and variance densities, \(\rho\) and \(\chi\). Consequently, the jammed occupation density is self-averaging.

The bulk coefficients nevertheless depend strongly on \(\lambda\) and on the particle composition. Charge-selective deposition generally increases the packing density by favoring geometrically efficient placements near gap boundaries. At \(\lambda=0.99\), the opposite-charge-dominated mixture reaches \(\rho=0.9451\) and reduces the variance density to \(\chi=0.0243\), whereas equal-charge dominance produces lower coverage and larger fluctuations. These results show that charge composition can control both the efficiency and the reproducibility of the jammed state.

\begin{acknowledgments} G. Palacios acknowledges fellowship support by the Fundação de Apoio ao Desenvolvimento da Universidade Federal de Pernambuco (FADE/UFPE), under Agreement No.~48/24 (FADE/UFPE/FINEP), Project No.~01.24.0540.00, Ref.~1020/24, associated with UFPE Agreement No.~34/2025 and administrative process No.~23076.095871/2024-80. A.M.S. Macêdo acknowledges financial support from CNPq (Grants 310800/2026-9 and 307626/2022-9) \end{acknowledgments}

\bibliography{references}

\end{document}